\documentclass[%
 reprint,
 amsmath,
 amssymb,
 aps,
 superscriptaddress,
 floatfix,
]{revtex4-1}

\usepackage{color}
\usepackage{siunitx}
\usepackage{graphicx}
\usepackage{braket}
\usepackage{verbatim}
\usepackage{dcolumn}
\usepackage{bm}
\usepackage{hyperref}
\usepackage[main=english,vietnamese]{babel}

\usepackage{accents}

\begin{document}

\preprint{APS/123-QED}

\def\mytitle{Kagom\'e network of chiral miniband-edge states in double-aligned graphene-hexagonal boron nitride structures}
\title{\mytitle{}}

\author{Christian Moulsdale}
\email{christian.moulsdale@postgrad.manchester.ac.uk}
\affiliation{\noindent School of Physics and Astronomy, University of Manchester, Manchester M13 9PL, UK}%
\affiliation{\noindent National Graphene Institute, University of Manchester, Manchester M13 9PL, UK}%

\author{\noindent Angelika Knothe}
\affiliation{\noindent National Graphene Institute, University of Manchester, Manchester M13 9PL, UK}%

\author{\noindent Vladimir Fal'ko}
\affiliation{\noindent School of Physics and Astronomy, University of Manchester, Manchester M13 9PL, UK}%
\affiliation{\noindent National Graphene Institute, University of Manchester, Manchester M13 9PL, UK}%
\affiliation{\noindent Henry Royce Institute, Institute for Advanced Materials, Manchester M13 9PL, UK}%

\date{\today}

\begin{abstract}
Twistronic heterostructures have recently emerged as a new class of quantum electronic materials with properties determined by the twist angle between the adjacent two-dimensional materials. Here we study moir\'e superlattice minibands in graphene (G) encapsulated in hexagonal boron nitride (hBN) with an almost perfect alignment with both the top and bottom hBN crystals. We show that, for such an orientation of the unit cells of the hBN layers that locally breaks inversion symmetry of the graphene lattice, the hBN/G/hBN structure features a Kagom\'e network of topologically protected chiral states with energies near the miniband edge, propagating along the lines separating the areas with different miniband Chern numbers. 
\end{abstract}

\maketitle

Recently, graphene-based systems have been shown to host various ``weak topological'' effects~\cite{PhysRevB.76.045302} among their electronic properties~\cite{Zhang2005,Novoselov2006,PhysRevLett.99.236809,Li2011,PhysRevB.88.121408,Ju2015,Yin2016,Li2016,PhysRevX.8.031023,PhysRevB.98.155435,PhysRevLett.121.257703,PhysRevLett.121.037702,doi:10.1126/sciadv.aaq0194,PhysRevB.98.035404,PhysRevLett.121.257702,Dutreix2019,PhysRevLett.124.126802,Shi2020,Li2020,doi:10.1021/acs.nanolett.0c04343}, which stem from the Berry phase/curvature in the electronic band structure of mononolayer graphene~\cite{PhysRevB.29.1685,doi:10.1143/JPSJ.67.2857} or its Bernal bilayers~\cite{PhysRevLett.96.086805,Novoselov2006,PhysRevB.84.041404,PhysRevB.101.085118,PhysRevB.98.155435,PhysRevLett.121.257702,PhysRevLett.124.126802,PhysRevB.101.235423,doi:10.1021/acs.nanolett.0c04343,2104.03399}. The topological effects manifest themselves in chiral states forming at the edges of the system or around internal structural defects, and propagating in opposite directions in the two valleys of graphene. These chiral states have been studied in detail in gapped bilayer graphene with either AB/BA domain boundaries~\cite{Zhang10546,Ju2015,Yin2016}, or an electrostatically inverted interlayer asymmetry gap~\cite{PhysRevLett.100.036804,Li2016}.

\begin{figure}[h!]
    \centering
    \includegraphics[width=\linewidth]{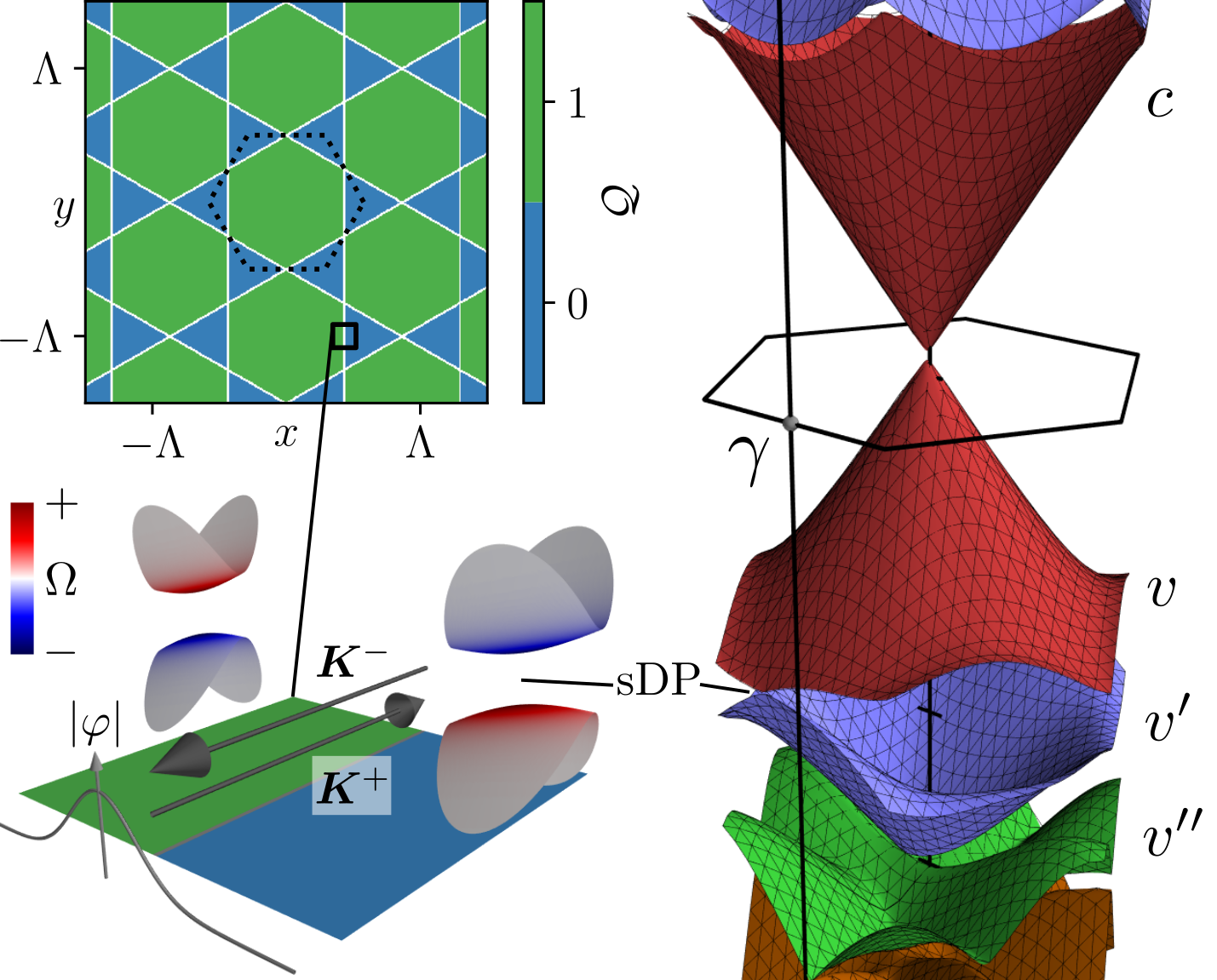}
    \caption{A typical miniband spectrum of graphene encapsulated into mutually aligned hBN crytals. \textit{Top inset.} Map of locally-defined $v$-miniband Chern number, $\mathcal {Q} $. \textit{Bottom inset.} Dispersion and Berry curvature at the $v$/$v'$ miniband edge in the gaped regions and chiral 1D modes counter-propagating in $K^\pm$ valleys along a Kagom\'e network of locally gapless $v$/$v'$ miniband edges. }
    \label{fig:1}
\end{figure}

Weak topological states have also emerged in the context of twistronic graphene systems~\cite{PhysRevB.88.121408,PhysRevLett.125.096402,PhysRevB.104.195410,Walet_2019,PhysRevLett.125.096402,Xu2019,PhysRevB.104.195410,PhysRevLett.121.037702,PhysRevB.98.035404} and in heterostructures of graphene (G) and hexagonal boron (hBN). The electronic properties of the latter system are qualitatively modified by the moir\'e superlattice~\cite{Yankowitz2012,PhysRevB.87.245408,doi:10.1126/science.aaf1095,Ren_2020} (mSL) 
with a period $\lambda\approx a/\sqrt{\delta^2 +\theta^2}$ (reaching $\SI {14} {nm} $ for small misalignment angles $\theta \rightarrow 0$, determined by the G-hBN lattice mismatch, $\delta \approx 0.018$). The system features a well-defined first miniband edge on the valence side of the graphene layer's dispersion~\cite{PhysRevB.87.245408,doi:10.1126/science.aaf1095}, as illustrated in Fig.~\ref{fig:1}.

The encapsulation of graphene between two hBN crystals with a high-precision alignment~\cite{doi:10.1126/sciadv.abd3655} leads to a further refinement of the superlattice effects, caused by the interference of Dirac electrons Bragg-scattered off the moir\'e superlattice (mSL) determined by the top and bottom G/hBN interfaces. Here, we study the influence of the relative lateral offset $\bm {\tau} $ between the top and bottom hBN crystals on moir\'e minibands in double-aligned hBN/G/hBN structures, considering the orientation (parallel versus antiparallel) of the unit cells in the hBN lattices. For graphene's minibands, the unit cell orientation matters due to the lack of inversion symmetry in the hBN monolayer, which is maximally passed onto graphene encapsulated between two hBN layers with parallel unit cell orientations but mutually cancels in the antiparallel case. 

\begin{figure}[h!]
    \centering
    \includegraphics[width=\linewidth]{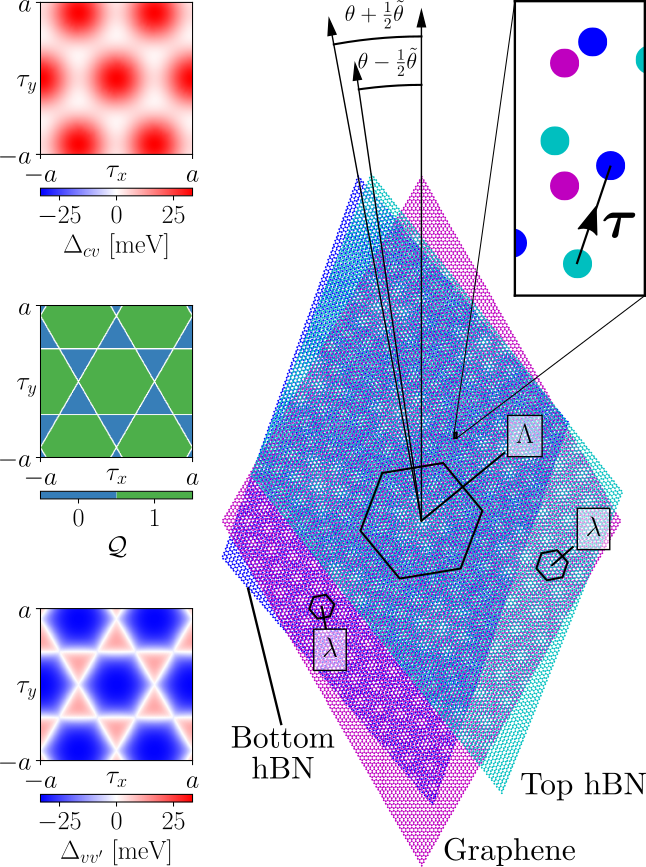}
    \caption{\textit {Right.} Graphene encapsulated between bottom and top hBN layers with twists $\theta \pm\frac {1} {2}\tilde {\theta} $ , respectively ($ |\tilde{\theta}| \ll \delta$). The interference of the layers results in a mSL of period $\lambda $, featuring a long-period variation of period $\Lambda $, whose unit cells are shown. \textit{Inset.} The offset vector $\bm{\tau}$ between the unit cells of the top and bottom hBN layers has components $(\tau_x, \tau_y) $ along the zigzag and armchair axes, respectively. \textit{Inset.} The valley Chern number $\mathcal {Q} $ of miniband $v$, the gap $\Delta_{cv}$ and the minigap $\Delta_{vv'}$ against offset for aligned hBN layers ($\tilde {\theta} = 0 $).}
    \label{fig:2}

\end{figure}

The inversion asymmetry, induced by hBN in graphene, leads to minigaps at the moir\'e miniband edges~\cite{PhysRevB.87.245408} (in particular, at the bottom edge of the first miniband on the valence band side, $ v $, corresponding to graphene doping of 4 holes per moir\'e supercell), of graphene's dispersion in Fig.~\ref{fig:1}, whose size, together with the Chern numbers of the minibands~\cite{doi:10.1143/JPSJ.74.1674,2016}, depends on the lateral offset between the top and bottom hBN crystals. For hBN/G/hBN structures with a small misalignment angle, $\tilde{\theta} \ll \delta $, between the top and bottom hBN layers, this offset varies across the coordinate space, as 
\begin{equation}
    \label{tau}
\bm{\tau} (\bm{r}) = \tilde{\theta} \bm{e}_z \times \bm{r}, 
\end{equation}
which leads to a long-period, $\Lambda\approx a/|\tilde {\theta} | $, variation of the mSL properties (see Fig.~\ref{fig:2}). A peculiar feature of this modulation is the closing and reopening of a minigap at the $v/v'$ miniband edge, which occurs along the lines forming a Kagom\'e structure in the real space, sketched in Fig.~\ref{fig:1}. Below, we study electron states at the $v/v'$ miniband edge, confined to this Kagom\'e network and discuss how chiral one-dimensional states (propagating in opposite directions in the $\bm {K} ^\pm $ valleys) provide this system with a finite conductivity even when its Fermi level would be set between the $v$ and $v'$ miniband edges in the gapped areas of the structure, with a characteristic pattern of Aharanov-Bohm oscillations. 

The above statement is based on the analysis of local miniband characteristics of the trilayer structure depicted in Fig.~\ref{fig:2}. Here, the bottom/top hBN monolayers with parallel orientation of their non-symmetric unit cells are twisted with respect to graphene by $\theta \pm\frac {1} {2}\tilde {\theta} $, respectively, with small mutual misalignment $|\tilde{\theta}| \ll \delta$, which determines the spatial variation of their local offset in Eq.~\ref{tau}. For each fixed offset $\bm{\tau}$, the Hamiltonian \cite{PhysRev.104.666,PhysRevB.87.245408,f474c4b4e81942508ccf4ca0eaeae7bf,DeltaAB} of electrons in the $\bm{K} ^\xi$ valley ($\xi=\pm$) of graphene is
\begin{align}
\label{eq:H}
\nonumber &\hat{H} =-i \hbar v \bm{\sigma} \cdot \nabla +
2 \sum_{\mathcal {P} = \pm} \sum_{m = 0}^5 \mathfrak{C}_m e^{i \bm{G}_m \cdot \bm{r}} U_{\mathcal {P}, m} + \frac{1}{2} \Delta_{cv} \sigma_z, \\
\nonumber & U_{\mathcal {P} , m} = u_0^\mathcal {P}\mathcal {P} ^{m + \frac{1}{2}} + (-\mathcal {P})^{m + \frac{1}{2}} (u_3^\mathcal {P} \sigma_z - i \xi u_1^\mathcal {P} \bm{e}_m \cdot \bm{\sigma})\\
\nonumber & \bm{\sigma} = (\xi \sigma_x, \sigma_y), \quad\Delta_{cv} \approx \frac{2}{3} \sum_{m = 0}^5 (\mathfrak{C}_m^2 \Delta_u - \mathfrak{S}_m^2 \Delta_h), \nonumber\\
&\mathfrak{C}_m = \cos (\tfrac {1} {2}\bm{g}_m \cdot \bm{\tau}), \quad\mathfrak{S}_m = \sin (\tfrac {1} {2}\bm{g}_m \cdot \bm{\tau}).
\end{align}
The values of the parameters used are given in Table S1 in the Supplemental Material (SM)~\cite{SM}, and the expression for graphene encapsulated between hBN monolayers with antiparallel orientation of their unit cells is given in SM S2~\cite{SM}.

The first term in Eq.~(\ref{eq:H}) is the Dirac Hamiltonian for electrons in monolayer graphene~\cite{PhysRev.104.666}. The second term describes the mSL produced by the  layers, with the reciprocal mSL vectors, $\bm{G}_m \approx \delta \bm{g}_m - \theta \bm{e}_z \times \bm{g}_m$, expressed in terms of the reciprocal lattice vectors of graphene, $\bm{g}_m = \frac{4 \pi}{\sqrt{3} a} \bm{e}_z \times \bm{e}_m, \quad \bm{e}_m = (\cos \frac{m \pi}{3}, \sin \frac{m \pi}{3})$, $m = 0, 1, ..., 5$. This term includes mSL potentials, sublattice asymmetry gaps and gauge fields of parity $\mathcal {P} = \pm$ (under spatial inversion), quantified using the parameters $ u_0 ^ \mathcal {P} $, $ u_3 ^ \mathcal {P} $ and $ u_1 ^ \mathcal {P} $, respectively.

The orientation and offset $\bm {\tau} $ of the unit cells in each hBN layer determine the magnitude of the odd-parity terms in Eq.~(\ref{eq:H}), which are responsible for the inversion symmetry breaking features in the dispersion, such as the opening of a minigap between minibands $v$ and $v'$. This corresponds to graphene doping of 4 holes per moir\'e supercell, with electron density $ -4 n_0 $ ($ n_0 = 2/\sqrt {3}\lambda ^ 2 $). In the parametrisation of Hamiltonian (\ref{eq:H}), we take into account that the positions of the graphene atoms rearrange to minimise graphene's adhesion energy with the hBN layers while maintaining the mSL period~\cite{PhysRevB.90.075428,PhysRevB.90.115152,Jung2015,DeltaAB}. The in-plane and out-of-plane rearrangements combine with the second term to give the respective contributions, $\Delta_u $ and $\Delta_h $, to the sublattice asymmetry gap $ \Delta_{cv} $. This gap appears in the odd parity, non-oscillatory third term, which breaks the sublattice symmetry of graphene. In antiparallel alignment, the contributions to the odd parity terms from each layer cancel, and inversion symmetry is preserved. Instead, we focus on parallel alignment where the inversion symmetry breaking is enhanced, depending strongly on the offset $\bm {\tau} $.

To study minibands of Dirac electrons in this system, we diagonalise Hamiltonian (\ref{eq:H}) using the basis of plane-wave Dirac states, folded onto the mSL Brillouin zone shown in Fig.~\ref{fig:1}. An example of a typical miniband dispersion is shown in Fig.~\ref{fig:1}, with other examples displayed in SM S3~\cite{SM}. Similarly to single-interface G/hBN heterostructures \cite{PhysRevB.87.245408,Yankowitz2012,Ponomarenko2013,Fuhrer2013,Yang2013,Yu2014,doi:10.1126/science.aaf1095}, this system features a well-defined first valence miniband, $ v $ for twists $ |\theta | \leq \SI{1}{\degree} $, whereas on the conduction band side the minibands strongly overlap on the energy axis. The inversion symmetry breaking produces a minigap, $\Delta_{vv'}$, at the edge between minibands $v$ and $v'$, whose magnitude, together with the $v/v'$ edge position in the Brillouin minizone, depends on the offset $\bm{\tau}$. The dependence of the minigap $\Delta_{vv'}$ on the offset $\bm{\tau}$ is shown in the bottom inset of Fig.~\ref{fig:2}. This panel shows that $\Delta_{vv'}$ (which is formally defined below) takes zero value and also changes sign on the lines which approximately correspond to the condition $\mathfrak{C}_m = 0$. This variation should be contrasted with the $\bm{\tau}$-dependence of the gap $\Delta_{cv}$ across the main Dirac point at the $c/v$ miniband edge shown on the top inset, where one can see that $\Delta_{cv}$ never changes sign. 

Along the lines on the $\bm{\tau}$ maps, where the minigap at the $v/v'$ miniband edge closes and reopens as a function of $\bm{\tau}$, the Chern number $ \mathcal{\xi Q} $~\cite{doi:10.1143/JPSJ.74.1674,2016} of miniband $v $ also changes (note that the miniband's Chern number has opposite sign in the $\bm{K}^\xi$ valleys, $\xi =\pm$). Here, $ \mathcal{Q} $ is found by computing the integral of the miniband’s Berry curvature over the mSL Brillouin minizone (see SM S4~\cite{SM} for details). The resulting map of $\mathcal{Q}(\bm{\tau})$ dependence is displayed as the middle inset in Fig.~\ref{fig:2}. The correlation between the behavior of the inversion-asymmetry gap $\Delta_{vv'}$ at its edge with miniband $ v' $ and of its Chern number suggests a simultaneous change of quantum  topological properties of states in both $v$ and $v'$, captured by the effective Hamiltonian \cite{PhysRevB.78.045415} applicable to the part of the Brillouin minizone in the vicinity of this edge,
\begin{align}
    \label{eq:H2}
    \nonumber H_{\bm{q}} ^ {vv'} = &\epsilon_{vv'} + \tfrac{1}{2} \Delta_{vv'} \sigma_z \\
    &+ \hbar (\xi v_s^x q_x\sigma_x + v_s^y q_y\sigma_y) +  \xi \hbar \bm{v}_a \cdot \bm{q},
\end{align}
whose basis is minibands $ v $ and $ v' $. Here, $\bm {q} $ is the wave vector relative to the position of the band edge, $\bm {v}_s $ and $\bm {v}_a $ are the symmetric and antisymmetric velocity, respectively, (the latter of which tilts the dispersion along the axis parallel to $\bm {v}_a $~\cite{PhysRevB.78.045415}) and $\epsilon_{vv'} $ is a constant energy shift. The parameters in Eq.~(\ref{eq:H2}) are fitted numerically to the minibands computed using Eq.~(\ref{eq:H}) SM S5~\cite{SM}. The sign of $\Delta_{vv'} $ is determined by the sign of the Berry curvature at the miniband edge, which changes simultaneously with the change of the Chern number. 

The variation of the offset $\bm{\tau} $ over the plane of a hBN/G/hBN structure, given by Eq.~(\ref{tau}), enables us to map the the computed dependence of miniband characteristics displayed on the insets in Fig.~\ref{fig:2} onto the real space: for this, we only need to rotate those plots by $\SI {90} {\degree} $ and rescale then by a factor $ 1/\tilde {\theta} $. This produces a Kagomé network of lines where the secondary minigap, $\Delta_{vv'}$, closes and then inverts its sign, and where the Chern number of miniband $ v $ changes (from 0 to 1). We show in SM S6~\cite{SM} that the shape of this network is independent of the model parameters. Topologically protected chiral channels form along these lines, supporting spin-degenerate, one-dimensional states which propagate in opposite directions in the time-reversed valleys.

The form and dispersion of these states can be found (in SM S7~\cite{SM}) by analyzing Hamiltonian (\ref{eq:H2}) with $\bm{q} \approx (q_{\parallel},-i\partial_{x_\perp})$ and $\Delta_{vv'} \approx x_\perp \partial_{x_\perp} \Delta_{vv'}|_{x_\perp=0}$ ($\partial_{x_\perp} \Delta_{vv'} |_{x_\perp=0} \sim | u_3 ^ - |/\Lambda >0 $) where $x_{\parallel}$ and $x_\perp$ are local coordinates along and perpendicular to the Kagom\'e network line. These states have a Jackiw-Rebbi \cite{PhysRevD.13.3398} form $\varphi_{ q_{\parallel}} \approx e^{i q_{\parallel} x_\parallel} e ^ {-x_\perp^ 2/2\aleph ^ 2}\zeta _{ q_{\parallel}}$ ($\zeta_{ q_{\parallel}}$ is a two-component vector), with Gaussian confinement within a length $\aleph\sim \sqrt {2\hbar v_s ^ x \Lambda/| u_3 ^ - |} $ perpendicular to the interface, and disperse linearly, $\epsilon(q_{\parallel}) = \xi \hbar \mathcal {V} q_{\parallel}$, with a 1D velocity $\mathcal {V}\sim v_s ^ x $. To consider these states independently  for each segment of the Kagom\'e network, we should require  $\lambda <\aleph\ll\Lambda $, which is satisfied for mismatches $|\tilde {\theta}| <\SI {0.1} {\degree} $.

\begin{figure}[h!]
    \centering
    \includegraphics[width=\linewidth]{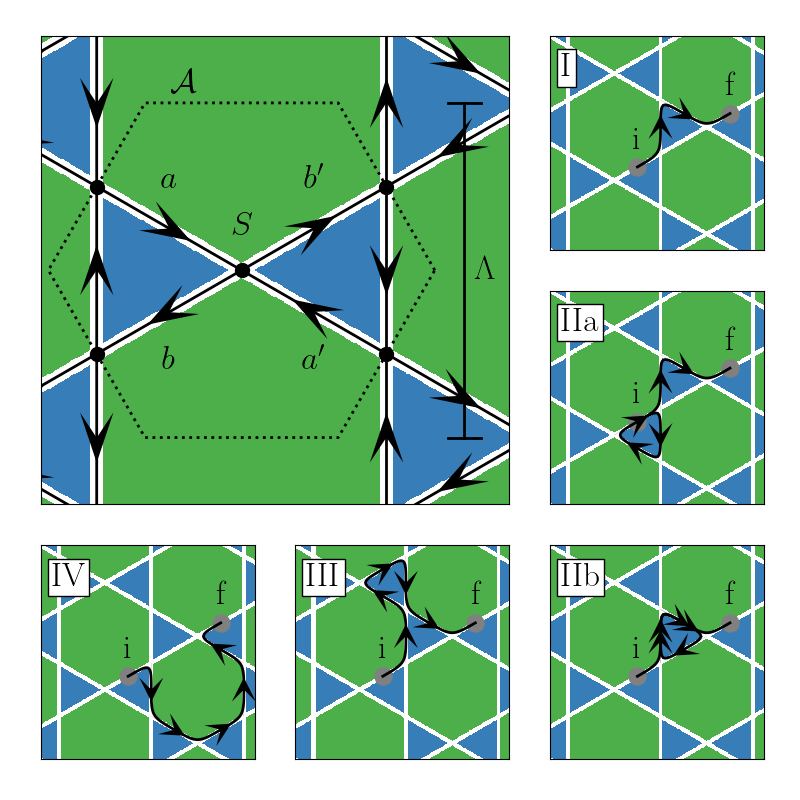}
    \caption{\textit{Top left.} The hexagonal structural element of the Kagomé network of chiral channels of area $\mathcal {A} =\sqrt{3}\Lambda ^ 2/2 $ and containing one $\mathcal {Q} = 0 $ bowtie and one $\mathcal {Q} = 1 $ hexagon. The chiral propagation of electrons in the $\bm{K}_+$ valley is shown, scattering at the three nodes. \textit{Clockwise from top right.} The five shortest paths for an electron wave packet to propagate from an injection position ``i'' to ``f'' (double arrows indicate a channel is traversed twice).}
    \label{fig:network}
\end{figure}

For a gapped moir\'e miniband spectrum, one could expect insulating behavior of a perfectly aligned hBN/G/hBN doped to the $v/v'$ miniband edge. For slightly ($|\tilde{\theta}|\ll \delta$) misaligned hBN crystals, the long-range variation of the local hBN-hBN offset and a network of chiral states, which it generates inside the minigap $\Delta_{vv'}$, quenches the resistivity of the hBN/G/hBN structure. While the exact calculations of the limiting resistivity would require a more rigorous consideration, based on the previous experiences of two-dimensional models for the network of 1D states~\cite{PhysRevLett.125.096402,PhysRevB.104.195410,PhysRevLett.126.186601}, we expect its value to correspond to a conductivity $\sigma_{xx}\sim e^2/h$ and to exhibit Aharonov-Bohm oscillations as a function of an out-of-plane magnetic field $B$. To describe the latter, we consider coherent electron waves (separately in the $\bm {K}_\pm $ valleys) on a network sketched in Fig.~\ref{fig:network}, where the structural element includes three nodes connected by chiral channels. At each node, an incoming wave packet scatters left or right according to the scattering matrix 
\begin{equation}
\label{eq: S}
\begin{pmatrix}
b\\b'
\end{pmatrix}
= S
\begin{pmatrix}
a\\a'
\end{pmatrix},\quad
S = e ^ {i\eta/3}
\begin{pmatrix}
\sqrt {P_R} & i\sqrt {P_L}\\
i\sqrt {P_L} &\sqrt {P_R}
\end{pmatrix}
\end{equation}
whose factor of $ i $ takes into account the Maslov's phase, and whose scattering probabilities, $P_R$ and $P_L$ ($ P_R+P_L=1$), can be considered as energy-independent within the narrow energy window of $\Delta_{vv'}$ (see SM S8~\cite{SM} for details). 

As a monochromatic wave of energy $\epsilon $ propagates across the network, its amplitude evolves according to the scattering matrix at the nodes and acquiring phase factors, $e^{i\epsilon\Lambda/2\mathcal {V}}$, after passing each ballistic segment of the network. At longer distances, partial waves, e.g., split from an incoming wave at ``i'' (see the side panels in Fig.~\ref{fig:network}), rejoin and interfere in another ballistic segment ``f''. An important feature of a periodic and $C_3$ rotationally symmetric network, such as in Figs.~\ref{fig:1} and \ref{fig:network}, is that the effect of the interference, constructive or destructive, of chiral edge states that travel from ``i'' to ``f'' along paths containing the same number, $N$, of segments does not depend on the exact energy (or wavelength) of the electron. This is because their ballistic phases, $e^{iN \epsilon\Lambda/2\mathcal {V}}$, are the same, producing the interference contribution determined only by their shapes through the energy-independent scattering amplitudes in Eq. (\ref{eq: S}). On the contrary, the interference of waves brought together by paths with a different number of segments, such as I and IV in Fig.~\ref{fig:network}), oscillates from constructive to destructive (and back) upon energy variation at the scale of $\hbar \mathcal {V}/\Lambda $.

Therefore, in the high-temperature regime, where $k_B T\gg \hbar \mathcal {V}/\Lambda $, the interference effects between waves arriving from ``i'' to ``f'' along paths of different lengths would be wiped out by the smearing of the Fermi step for electrons. The interference between waves brought from ``i'' to ``f'' by same-length paths (such as IIa, IIb, III and IV in Fig. \ref{fig:network}) would survive thermal averaging, without suppression, though this contribution would be sensitive to the external magnetic field, due to the Aharonov-Bohm phases from magnetic field fluxes encircled by the pairs of same-length paths. 

When discussing the interference effects in electronic transport at high temperatures, we are also conscious of the inelastic decoherence of electron waves, which efficiently destroys interference effects for the longer paths. For a system with decoherence length $\ell$, this can be accounted by a suppression factor $e^{-\Lambda/4\ell}$ applied to each ballistic segment of the Kagom\'e network. Therefore, to discuss the high-temperature limit, we consider the shortest paths that can contribute to the interference effect in transport, shown in the side panels of Fig. \ref{fig:network}. These paths are related to the `forward' electron propagation from a segment in one network unit cell to the equivalent segment in the next one, counted in the direction of the propagation of the chiral edge state (for valley K). These shortest paths contain three (I) and six (IIa, IIb, III, IV) ballistic segments of length $\Lambda/2 $, and the interfering amplitude at the point ``f'' for a wave starting at ``i'' with unit amplitude would be
\begin{align*}
\psi \approx  &
- e ^ {i \pi \phi / 4 \phi_0} \sqrt{P_L ^ 2 P_R} z ^ 3 
+ z ^ 6 e ^ {i \pi \phi / 2 \phi_0} \times\\ 
& \qquad  \qquad (- 2 P_L P_R ^ 2 
+ P_L ^ 2 P_R 
+ e ^ {-i 2\pi \phi / \phi_0} P_L ^ 2 P_R ) ,
\end{align*}
where $z = e ^ {i \eta / 3} e ^ {i\epsilon\Lambda/2\hbar \mathcal {V}} e ^ {-\Lambda / 4 \ell} $. We also account for additional phases, induced by the out-of-plane magnetic field and described in terms of magnetic field flux, $\phi = B\mathcal {A}$ through the unit cell area, $\mathcal {A} =\sqrt {3}\lambda ^ 2/2 $, of the Kagom\'e network ($\phi_0=h/e$ is the flux quantum). 

Then, we express the probability $\langle \mathcal {W} \rangle_{T}$ for the electron to get from the segment ``i'' to ``f'', averaged over the $k_BT$ energy interval near the Fermi level, as
\begin{align}
\label {eq:W}
\nonumber\langle \mathcal {W} \rangle_{T} \approx &
P_L ^ 2P_R |z | ^ 6
+ [4P_L ^ 2P_R ^ 4 + 2P_L ^ 4P_R ^ 2]|z | ^ {12}\\
&  + 2 [P_L ^ 4P_R ^ 2 - 2P_L ^ 3P_R ^ 3] |z | ^ {12} \cos \bigg (\frac{2 \pi \phi}{\phi_0}\bigg),
\end{align}
whose second line originates from the encircled Aharonov-Bohm phases, which, for the pairs of shortest paths, are all determined by the magnetic field flux through the unit cell area of the Kagom\'e network. As the probability, described by Eq.~(\ref{eq:W}), is a characteristic of the forward propagation of electrons, its oscillations also determine the Aharonov-Bohm oscillations in the network conductivity (see SM S9~\cite{SM} for backwards propagation), 
\begin{equation}
\label{eq:sigma}
\sigma_{xx} (\phi) \approx \frac{e^2}{h} \bigg [ \alpha +\beta e ^ {-5\Lambda/2\ell}\cos \bigg (\frac{2 \pi \phi}{\phi_0}\bigg) \bigg],
\end{equation}
where $\alpha,\beta\sim 1 $.

Overall, we have demonstrated the existence of a Kagom\'e network of chiral states lying in the minigap at the edge of the first moir\'e miniband on the valence band side of graphene encapsulated between hBN with parallel unit cells. This edge state network gives rise to quenched resistivity, $\sim h/e^2$, of graphene even when its Fermi level doping reaches that minigap. This conductivity, in Eq. (\ref{eq:sigma}), exhibits Aharonov-Bohm oscillations, whose period is determined by the area of the unit cell of the Kagom\'e network and, consequently, the misalignment. For the networks with a longer decay length $\ell $ (or a shorter period), the magnitude of the Aharonov-Bohm oscillations should increase, accompanied by the emergence of a finer structure, composed of  higher frequency harmonics corresponding to rational factor between the magnetic field flux through the Kagom\'e network cell and flux quantum, $\phi_0$. Such edge state networks emphasize the role of twistronic heterostructures as hosts of  topological phenomena and deserve further theoretical studies, e.g., taking into account electron-electron interactions in the chiral channels \cite{PhysRevLett.104.216406}.

We thank S. Slizovskiy and K. Novoselov for useful discussions. We acknowledge support from EU Graphene Flagship Project, EPSRC Grants No. EPSRC CDT Graphene-NOWNANO EP/L01548X/1, No. EP/S019367/1, No. EP/P026850/1, and No. EP/N010345/1, and EC Quantum Flagship Project No. 2D-SIPC.

All the research data supporting this publication are directly available within this publication and Supplemental Material accompanying this publication.

\bibliography{ref}

\end{document}


\def\mytitle{Kagom\'e network of chiral miniband-edge states in double-aligned graphene-hexagonal boron nitride structures}
\title{Supplemental material for ``\mytitle{}''}

\author{Christian Moulsdale}
\email{christian.moulsdale@postgrad.manchester.ac.uk}
\affiliation{\noindent School of Physics and Astronomy, University of Manchester, Manchester M13 9PL, UK}%
\affiliation{\noindent National Graphene Institute, University of Manchester, Manchester M13 9PL, UK}%

\author{\noindent Angelika Knothe}
\affiliation{\noindent National Graphene Institute, University of Manchester, Manchester M13 9PL, UK}%

\author{\noindent Vladimir Fal'ko}
\affiliation{\noindent School of Physics and Astronomy, University of Manchester, Manchester M13 9PL, UK}%
\affiliation{\noindent National Graphene Institute, University of Manchester, Manchester M13 9PL, UK}%
\affiliation{\noindent Henry Royce Institute, Institute for Advanced Materials, Manchester M13 9PL, UK}%

\date{\today}

\maketitle
\onecolumngrid

\section {Parameters used in hBN/G/hBN Hamiltonian in Eq. (2)}
\label{sec: parameters}

\begin{table}[h]
\centering
\begin{tabular}{|c|c|c|}
\hline
Parameter & Value & Reference \\
\hline\hline
$ a $ & $\SI {0.246} {nm} $ & \cite{RevModPhys.81.109}\\
\hline
$ v $ & $\SI {e6} {m/s} $ & \cite{RevModPhys.81.109} \\
\hline
$ \delta $ & $ 0.018 $ & \cite{PhysRevB.87.245408} \\
\hline
$ u_0 ^ \pm $ & $\pm V_\pm /2 $ & \cite{PhysRevB.87.245408} \\
\hline
$ u_1 ^ \pm $ & $-V_\pm $ & \cite{PhysRevB.87.245408} \\
\hline
$ u_3 ^ \pm $ & $-\sqrt{3}V_\pm /2 $ & \cite{PhysRevB.87.245408} \\
\hline
$ V_+ $ & $\SI {17} {meV} $ & \cite{doi:10.1126/science.aaf1095} \\
\hline
$ V_- $ & $\SI {3} {meV} $ & \cite{doi:10.1126/science.aaf1095} \\
\hline
$ \Delta_u $ & $\SI {8} {meV} $ & \cite{DeltaAB} \\
\hline
$ \Delta_h $ & $\SI {2} {meV} $ & \cite{DeltaAB} \\
\hline
\end{tabular}
\caption{Table of parameters used in Eq.~(2) of the main text.}
\label{tab:parameters}
\end{table}

The table of parameters used in Eq.~(2) of the main text is shown in Tab.~\ref{tab:parameters}. An alternative parameterisation of the superlattice interaction is found in~\cite{PhysRevB.87.245408}. The alternative superlattice potentials $ u_{0, 1, 2, 3} $ and $ \tilde {u}_{0, 1, 2, 3} $ are related to the potentials $ u_{0, 1, 3} ^\pm $ in this paper by
\begin{align}
\nonumber & u_0 = u_0 ^ +,\quad
u_1 = \frac{\delta}{\sqrt{\delta ^ 2+\theta ^ 2}} u_1 ^ +,\quad
u_2 = \frac{-\theta}{\sqrt{\delta ^ 2+\theta ^ 2}} u_1 ^ +,\quad
u_3 = u_3 ^ +, \\
& \tilde {u}_0 = u_0 ^ -,\quad
\tilde {u}_1 = \frac{\delta}{\sqrt{\delta ^ 2+\theta ^ 2}} u_1 ^ -,\quad
\tilde {u}_2 = \frac{-\theta}{\sqrt{\delta ^ 2+\theta ^ 2}} u_1 ^ -,\quad
\tilde {u}_3 = u_3 ^ -.
\end{align}

\section {Hamiltonian for antiparallel orientation of the hBN unit cells}
\label{sec:Hamiltonian}

The Hamiltonian of graphene encapsulated between two aligned layers of hBN depends on the alignment of the unit cells in the hBN layers. The Hamiltonian for parallel unit cells is given in Eq.~(2) of the main text, featuring odd parity terms that break inversion symmetry. The Hamiltonian for antiparallel unit cells is,
\begin{align}
\label{eq:HA}
\nonumber & \textstyle \hat{H} =
-i \hbar v \bm{\sigma} \cdot \nabla + 2 \sum_{m = 0}^5 e^{i \bm{G}_m \cdot \bm{r}}\times\\
\nonumber &\textstyle\quad\quad\quad\quad \{[\mathfrak{C}_m u_0^+ - (-1) ^ m \mathfrak{S}_m u_0 ^ -] + i[(-1) ^ m \mathfrak{C}_m u_1^+ + \mathfrak{S}_m u_1 ^ -] \sigma_z +\xi [(-1) ^ m \mathfrak{C}_m u_3^+ + \mathfrak{S}_m u_3 ^ -] \bm{e}_m \cdot \bm{\sigma})\}, \\
\nonumber & \textstyle \bm{\sigma} = (\xi \sigma_x, \sigma_y), \quad
\mathfrak{C}_m = \cos (\bm{g}_m \cdot \bm{\tau} / 2), \quad
\mathfrak{S}_m = \sin (\bm{g}_m \cdot \bm{\tau} / 2) \\
& \textstyle \bm{G}_m \approx \delta \bm{g}_m - \theta \bm{e}_z \times \bm{g}_m, \quad \bm{g}_m = \frac{4 \pi}{\sqrt{3} a} \bm{e}_z \times \bm{e}_m, \quad \bm{e}_m = (\cos \frac{m \pi}{3}, \sin \frac{m \pi}{3}),\quad
m = 0, 1, ..., 5,
\end{align}
where $\bm {\tau} $ is the lateral offset between the unit cell centres in each hBN layer. Note that the odd parity superlattice potentials $ u_{0, 1, 3} ^ - $ result in a contribution to the Hamiltonian which has even parity under inversion and inversion symmetry is preserved in this system.

\section {Examples of calculated miniband structures }
\label{sec:minibands}

\begin{figure}[h]
    \centering
    \includegraphics[width=\linewidth]{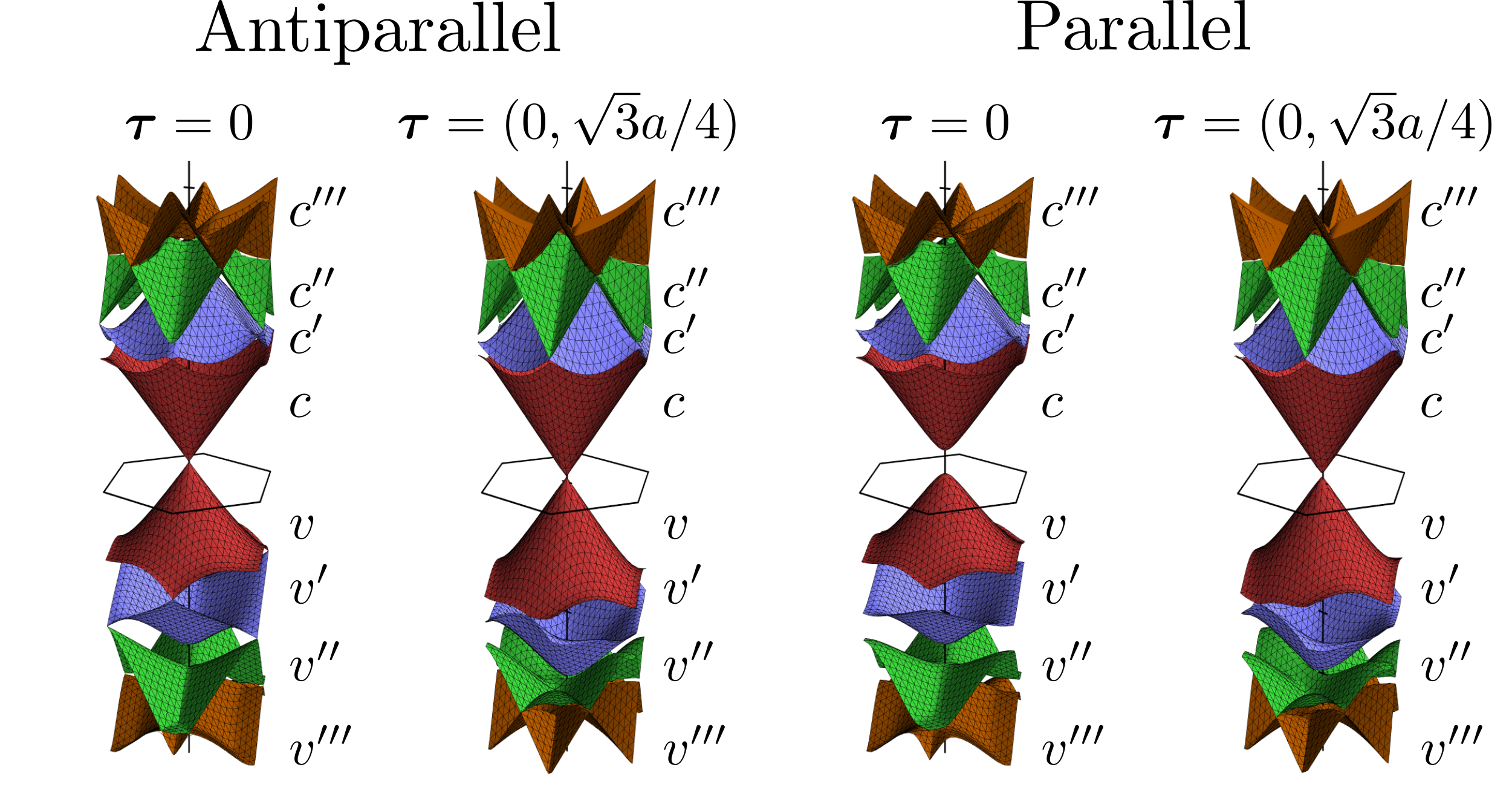}
    \caption{Miniband dispersions for parallel and antiparallel alignment of the unit cells in the hBN layers with various lateral offsets $ \bm{\tau } $ between the hBN layers.}
    \label{fig:bands}
\end{figure}

The first term of the Hamiltonians in Eq.~(2) of the main text and (\ref{eq:HA}) is the Dirac Hamiltonian of electrons in graphene. This has plane wave solutions $\psi_{\bm{k}}^ {c_0/v_0} (\bm{r}) =\frac {1} {\sqrt {2}} (1, \pm e ^ {i \xi \arctan (k_y/k_x)}) $, corresponding to the conduction $ (c_0) $ and valence $ (v_0) $ bands of energy $ \pm \hbar v |\bm{k} | $, respectively. We diagonalise the full Hamiltonians using the zone-folded wavefunction $\Psi_{\bm{k}} (\bm{r}) = \sum_{\alpha = c_0, v_0} \sum_{\bm{G}} c_{\bm{k} + \bm{G}}^\alpha \psi_{\bm{k} + \bm{G}}^\alpha (\bm{r})$ of wave vector $\bm{k} $ in the moir\'e superlattice (mSL) Brillouin minizone (sBZ). In the space spanned by the coefficients $\{c_{\bm{k} + \bm{G}}^\alpha\} $, the Hamiltonian is represented by a $2M \times 2M$ matrix $\mathcal{H}_{\bm{k}}$. Terms in Eq.~(2) of the main text and (\ref{eq:HA}) containing $e^{i \bm{G}_m \cdot \bm{r}} $ give off-diagonal elements in $\mathcal{H}_{\bm{k}}$ between coefficients with momenta separated by $ \bm{G}_m $. We sum over the shortest $ M $ reciprocal mSL vectors $\bm {G} $ required for the lowest four energy eigenvalues of $\mathcal{H}_{\bm{k}}$ to converge. In this case, convergence is reached with the $ M = 19 $ vectors which are the sum of at most two of the six shortest non-zero reciprocal mSL vectors $\bm{G}_m $ $ (m = 0, 1, ..., 5 ) $.

The conduction and valence band reconstruct into $2M$ minibands of energy $\epsilon_{n \bm{k}}$ and wavefunction $\chi_{n \bm{k}}$ in the coefficient basis. We show the dispersion of the first four valence $ (v, v', v'', v''')$ and conduction $ (c, c', c'', c''')$ minibands for zero offset $ (\bm{\tau} = 0) $ in Fig.~\ref{fig:bands}. The minibands are similar for each alignment, except antiparallel alignment of the hBN unit cells, which preserves inversion symmetry, gives a gapless spectrum and parallel alignment, which breaks inversion symmetry, gives a gapful spectrum. We also show the dispersions for an offset $\bm{\tau} = (0, \sqrt{3} a/4) $ satisfying $\mathfrak {C}_0 = 0 $. In this case, despite the inversion symmetry breaking, there are band closures in the dispersion for parallel alignment, including the secondary minigap on the valence side between minibands $ v $ and $ v' $ (as discussed in the main text). 

\begin{figure}[h]
    \centering
    \includegraphics[width=\linewidth]{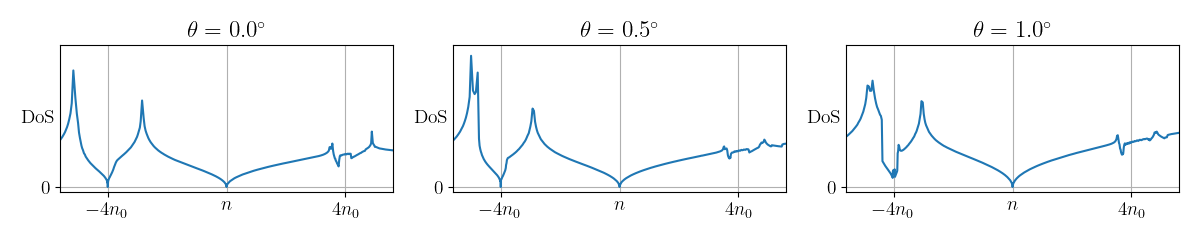}
    \caption{The density of states against electron density $ n $ for various twists with offset $ \bm{\tau } = (0, \sqrt{3} a/4) $.}
    \label{fig:dos}
\end{figure}

We show the density of states for parallel hBN unit cells against the electron density $ n $ (including the four-fold spin and valley degeneracy) for the offset $\bm{\tau} = (0, \sqrt{3} a/4) $ which closes the secondary minigap in Fig.~\ref{fig:dos}. A filled miniband contributes a value $ n_0 = 2/(\sqrt{3}\lambda ^ 2) $ to the electron density, so the edge between the minibands $ v $ and $ v' $ corresponds to electron density $ n = - 4n_0 $. For small twists, $|\theta| < \SI{1}{\degree} $, the density of states is zero at $ n = - 4n_0 $, and there is no overlap between these bands on the energy axis. There is overlap on the conduction side between minibands $ c $ and $ c' $ since the density of states is non-zero at $ n = 4n_0 $. For larger twists, $|\theta| \geq \SI{1}{\degree} $, minibands $ v $ and  $ v' $ overlap with non-zero density of states at $ n = -4n_0 $.

\section {Topological characteristics of the minibands }
\label{sec:topology}

\begin{figure}[h]
    \centering
    \includegraphics[width=\linewidth]{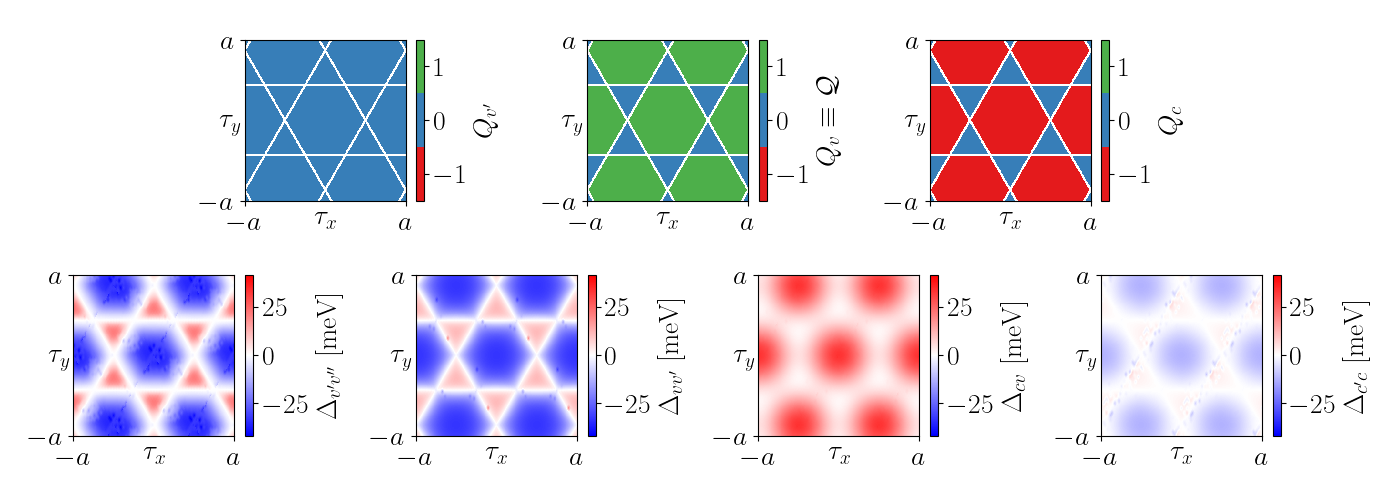}
    \caption{\textit {Top row.} Valley Chern numbers of minibands $v'$ to $ c $ against offset $\bm{\tau}$. \textit {Bottom row.} The minigaps $\Delta_{v'v''} $, $\Delta_{vv'} $, $\Delta_{cv} $ and $\Delta_{c'c} $ of the miniband edges $ v'/v'' $, $ v/v' $, $ c/ v$ and $ c'/c $.}
    \label{fig:Delta}
\end{figure}

When inversion symmetry is broken, the topology of miniband $ n $ is represented by its valley Chern number $\xi Q_n $~\cite{PhysRevLett.49.405,1984,KOHMOTO1985343,doi:10.1143/JPSJ.74.1674,PhysRevLett.99.236809,RevModPhys.82.1959,2016}, which is given by the integral of the miniband's Berry curvature $\Omega_{n \bm{k}} $ over the mSL Brillouin minizone,
\begin{align}
\label{eq:C}
\nonumber & Q_n = \frac{1}{2 \pi} \int_\mathrm{sBZ} d^2 k \, \Omega_{n \bm{k}}, \\
& \Omega_{n \bm{k}} = -2 \, \mathrm{Im} \sum_{m \neq n} \frac{[\chi_{n \bm{k}} ^\dag (\partial_{k_x} \mathcal{H}_{\bm{k}}) \chi_{m \bm{k}}] [\chi_{m \bm{k}} ^\dag (\partial_{k_y} \mathcal{H}_{\bm{k}}) \chi_{n \bm{k}}]}{(\epsilon_{n \bm{k}} - \epsilon_{m \bm{k}})^2}.
\end{align}
In the expression for the Berry curvature, we sum across the other minibands $m$. The valley Chern number has opposite signs in each valley since the Berry curvature has odd parity under inversion as a result of time reversal symmetry. In the main text, we define the valley Chern number of the first valence miniband as $\mathcal {Q} \equiv Q_v $. We plot the valley Chern numbers of the second valence miniband $v'$ to the first conduction miniband $ c $ against offset $\bm{\tau}$ in Fig.~\ref{fig:Delta}. The valley Chern numbers of minibands $ v $ and $ c $ change as the $\mathfrak {C}_m = 0 $ lines are crossed in offset space and the secondary minigaps close.

\section {Effective Hamiltonian at the $ v/v' $ miniband edge}
\label{sec: effective}

The secondary minigap on the valence side, $\Delta_{vv'} $, closes for offsets along the $\mathfrak {C}_m = 0 $ lines. For offsets near these lines $ (\mathfrak {C}_m\sim 0 ) $, the minigap is non-zero at a miniband edge appearing at wavenumber $\bm{k}_e $. Expanding about the miniband edge, $\bm{q} =\bm{k} - \bm{k}_e $, the first and second valence minibands have dispersions~\cite{PhysRevB.78.045415},
\begin{equation}
\epsilon_{ v/v'}\approx \epsilon_{vv'} + \xi \hbar \bm{v}_a \cdot \bm{q} \pm \sqrt{(\hbar v_s^x q_x) ^ 2 + (\hbar v_s^y q_y) ^ 2 + (\Delta_{vv'}/2) ^ 2}.
\end{equation}
This is parameterised by a constant energy shift $\epsilon_{vv'} $, symmetric velocity $\bm{v}_s $, anti-symmetric velocity $\bm{v}_a $ and the secondary minigap $\Delta_{vv'} $, whose sign is determined by the Berry curvature,
\begin{equation}
\Omega_{ v/v'} \approx \mp \frac{1}{4} \xi \Delta_{vv'} v_s^x v_s^y [(\Delta_{vv'}/2) ^ 2+ (\bm{v}_s \cdot \bm{q}) ^ 2] ^ {-3/2}.
\end{equation}
The miniband edge does not appear at a high symmetry point of the mSL Brillouin minizone since the offset is non-zero. Hence, the anti-symmetric velocity is non-zero $ (\bm{v}_{\mathrm{a}} \neq 0) $, and the dispersion is tilted~\cite{PhysRevB.78.045415}. The dispersion and Berry curvature are described by the effective Hamiltonian in Eq.~(3) of the main text and are plotted in Fig.~1 of the main text. We plot the secondary minigap $ \Delta_{vv'} $ against offset $\bm {\tau} $ in Fig.~\ref{fig:Delta}. As described in the main text, the sign of the secondary minigap $ \Delta_{vv'} $ changes sign as the $\mathfrak{C}_m = 0$ lines are crossed, inverting the Berry curvature of the minibands. This transfers a unit of valley Chern number between the minibands, giving the transition in $\mathcal {Q} $ in Fig.~\ref{fig:Delta}. 

We can derive a similar effective Hamiltonian at the edge between the first conduction miniband $ c $ and the second conduction miniband $ c' $. This allows us to define the secondary minigap on the conduction side, $\Delta_{c'c} $, which we plot in Fig.~\ref{fig:Delta}. The minibands undergo a similar inversion at this band edge as the $\mathfrak {C}_m = 0 $ lines are crossed, transferring a unit of valley Chern number between them. Finally, we repeat this process for the edge between the second and third valence minibands, $ v '$ and $ v ''$ respectively, with the corresponding minigap $\Delta_{v'v''} $ also changing sign as the $\mathfrak {C}_m = 0 $ lines are crossed. The inversion of the miniband structure at the edges of the second valence miniband with the first and third valence minibands cancels out pairwise and $ Q_{ v'}\equiv 0 $. The principal gap $\Delta_{cv} $ between the first valence miniband and the first conduction miniband never closes.

\section {Robustness of the Kagom\'e network of chiral states}
\label{sec: network}

\begin{figure}[h]
    \centering
    \includegraphics[width=\linewidth]{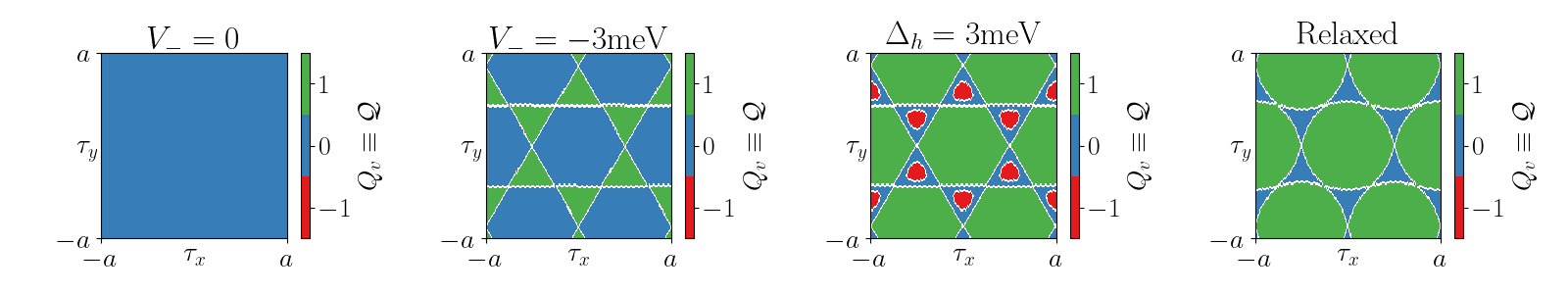}
    \caption{The valley Chern number $\mathcal{Q}$ of the first valence miniband $ v $ against offset $(\tau_x, \tau_y)$ for various modifications to the parameters in Tab.~\ref{tab:parameters}.}
    \label{fig:dif}
\end{figure}

In Fig.~\ref{fig:dif}., we demonstrate that our main finding of a Kagomé network of topologically protected one-dimensional chiral channels is robust against the choice of model parameters within a reasonable range. The phase transitions of the valley Chern number $ \mathcal {Q} $ of the first valence miniband $ v $ are associated with the change in sign of the secondary minigap $ \Delta_{vv'} $ at its edge with the second valence miniband $ v' $. Hence, flipping the sign of the odd parity superlattice potential $V_-$ swaps the phases and reverses the direction of the chiral channels while preserving the network. Turning off the inversion symmetry breaking in the interaction with the hBN layers by setting $V_- = 0$ prevents the transitions entirely as seen when comparing Fig.~2 of the main text to Fig.~\ref{fig:dif}, and there is no network of channels.

If $\Delta_h / \Delta_u > \frac{1}{3}$, then the principal gap $\Delta_{cv}$ in Eq.~(2) of the main text closes near the offset unit cell corners, giving the second phase transition to $\mathcal {Q} = -1 $ observed in Fig.~\ref{fig:dif} for $\Delta_h = \SI{3}{meV}$. This would produce a disconnected network of channels in the principal gap, which would not affect the network at the secondary minigap. A full treatment of the relaxation of the graphene layer onto the hBN layers results in terms other than $\Delta_{cv}$~\cite{DeltaAB}. The most significant of these mixes the superlattice potentials $u_j^s$, suppressing the odd potentials $u_j^-$. With an increased $V_- = \SI{6}{meV}$ to compensate, we find the phase diagram in Fig.~\ref{fig:dif}. This forms a Kagomé network which is qualitatively similar to Fig.~2 of the main text.

\section {Chiral states at the $ v/v' $ miniband edge }
\label{sec:states}

We consider interfaces along the $ y $-axis in Fig.~1 of the main text, corresponding to the $\mathcal {C}_0 = 0 $ lines in offset space. The other interfaces are related by the six-fold symmetry of the network. At the edge between the first and second valence minibands, the fitted parameters in Eq.~(3) are $ v_s ^ x \approx 0.2 v $, $ v_s ^ y \approx 0.9 v $, $ v_a ^ x \approx \pm0.1 v $ (for interfaces along the $\mathcal {Q} = 0 $ triangles of shape $\triangleright $ and $\triangleleft $, respectively) and $ v_a ^ y = 0 $. We choose the basis with components perpendicular and parallel to the interface with unit vectors $\bm {e}_\perp =\pm\bm {e}_x $ and $\bm {e}_\parallel =\pm\bm {e}_y $, respectively, such that the gradient of the secondary minigap perpendicular to the interface is $\partial_{x_\perp} \Delta_{vv'} \approx \SI[parse-numbers = false] { 0.7 |\tilde {\theta} |} {eV / nm} >0 $.

After performing the substitution $\bm{q} \rightarrow (q_{\parallel},-i\partial_{x_\perp})$ discussed in the main text, we solve with the Jackiw-Rebbi ansatz $\varphi_{q_\parallel} = e^{i {q_\parallel} x_\parallel} \mathcal {J}_{q_\parallel} (x_\perp) $ for the wavefunction of the channel state which propagates with wavenumber $ {q_\parallel} $ along the interface~\cite{PhysRevD.13.3398}. We use the finite difference method to find the two-component vector function $\mathcal {J}_{q_\parallel} (x_\perp) $ which confines the state perpendicular to the interface, subject to the boundary conditions $\lim_{x_\perp\rightarrow\pm\infty}\mathcal {J}_{q_\parallel} (x_\perp) = 0 $. This takes the Gaussian form $\mathcal {J}_{q_\parallel} (x_\perp) \approx e ^ {-x_\perp/2\aleph ^ 2}\zeta_{q_\parallel} $, where $\aleph \approx 2\tilde {\theta} ^ {- 1/2 } a $ and $\zeta_{q_\parallel} $ is a two-component vector which is independent of position. The state has linear dispersion $\xi \hbar\mathcal {V} {q_\parallel} $, with velocity $\mathcal {V}\approx 0.3 v $. The direction of propagation is opposite for states along the triangles $\triangleright $ and $\triangleleft $ and for each valley. The channel states and their confinement are shown in Fig.~1 of the main text. We require that the lateral confinement of the channel state is longer than the mSL period ($\aleph >\lambda $) and much shorter than the period of the long-range variation ($\aleph\ll\Lambda $). These give the constraints $|\tilde {\theta}| <4\delta ^ 2 $ and $|\tilde {\theta}|\ll 1/4 $, respectively. We assume that the mismatch is small, $|\tilde {\theta}|\ll\delta $, when we construct the system, and the latter constraint is automatically met. Hence, the only constraint is that $ |\tilde {\theta} | <\SI {0.1} {\degree} $. 

\section {Kagomé network sites and scattering amplitudes for chiral states}
\label{sec:scattering}

\begin{figure}[h!]
    \centering
    \includegraphics[width=\linewidth]{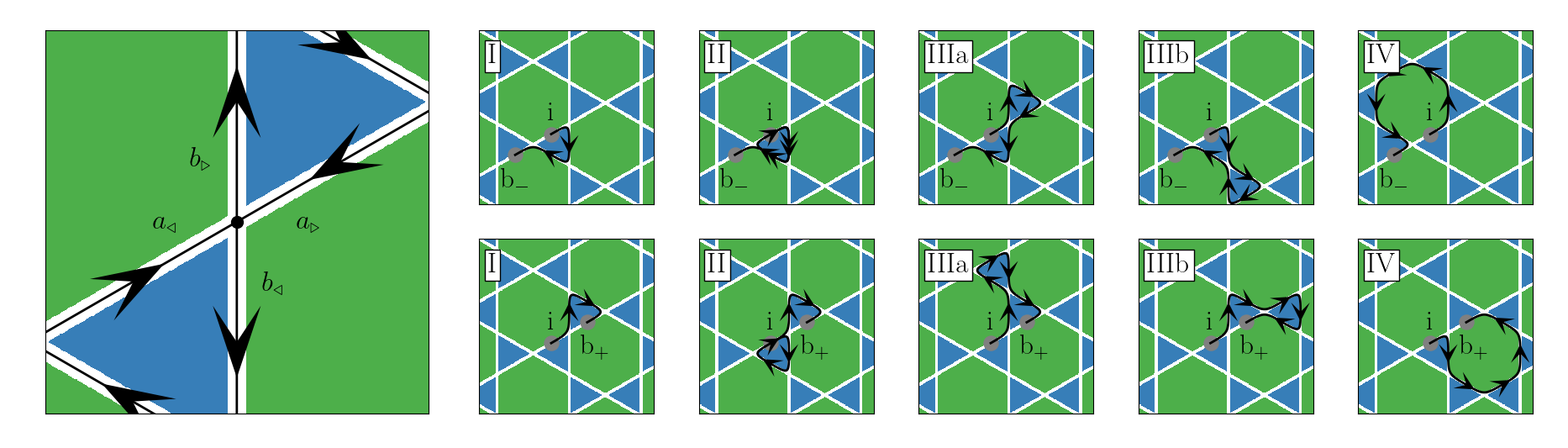}
    \caption{\textit{Left.} Scattering of channel states in the $\bm {K} ^ + $ valley at a node of the network. The amplitudes of the incoming $ (a_{\triangleright/\triangleleft}) $ ($ a/a' $ in the main text, respectively) and outgoing $ (b_{\triangleright/\triangleleft}) $ ($ b/b' $ in the main text, respectively) modes along the $ \mathcal{Q} = 0 $ triangles $\triangleright $ and $\triangleleft $, respectively, are highlighted. \textit{Right.} The five shortest paths for an electron wave packet in the $\bm {K} ^ + $ valley to propagate from an injection position ``in'' to ``$\mathrm{b}_\mp$'' (\textit {Top and bottom rows respectively.}). Double arrows indicate a channel is traversed twice.}
    \label{fig:scattering}
\end{figure}

We show scattering of channel states in the $\bm {K} ^ + $ valley at a node of the network in Fig.~\ref{fig:scattering} (we neglect scattering with states in the $\bm {K} ^ - $ valley). The amplitudes of the incoming $ (a_{\triangleright/\triangleleft}) $ ($ a/a' $ in the main text, respectively) and outgoing $ (b_{\triangleright/\triangleleft}) $ ($ b/b' $ in the main text, respectively) states on channels along the $ \mathcal{Q} = 0 $ triangles $\triangleright $ and $\triangleleft $, respectively, are related by the scattering matrix $ S $ of the node:
\begin{equation}
\label{eq:S}
\begin{pmatrix}
b_\triangleright \\
b_\triangleleft
\end{pmatrix}
= S
\begin{pmatrix}
a_\triangleright \\
a_\triangleleft
\end{pmatrix}, \quad
S =
\begin{pmatrix}
S_{\triangleright,\triangleright} & S_{\triangleright,\triangleleft}\\
S_{\triangleleft,\triangleright} & S_{\triangleleft,\triangleleft}
\end{pmatrix}.
\end{equation}
We constrain the scattering amplitudes using the symmetries of the system:
\begin{itemize}
    \item Unitarity: $ (S ^ {-1} = S ^\dag) $
    \item Time reversal symmetry: $ (S ^ T = S) $
    \item Six-fold rotational symmetry:
\begin{itemize}
    \item $\SI {120} {\degree} $ rotation: $ S $ is the same at each node
    \item $\SI {180} {\degree} $ rotation: $ S_{\triangleright, \triangleright} = S_{\triangleleft, \triangleleft}$ and $ S_{\triangleright, \triangleleft} = S_{\triangleleft, \triangleright}$
\end{itemize}
\end{itemize}
This gives the scattering amplitudes $ S_{\triangleright, \triangleright} = S_{\triangleleft, \triangleleft} = e ^ {i \eta/3}\sqrt {P_R} $ and $ S_{\triangleright, \triangleleft} = S_{\triangleleft, \triangleright} = i e ^ {i \eta/3}\sqrt {P_L} $ described in the main text, with real parameters $\eta $, $ P_L $ and $ P_R = 1 - P_L $ (the latter of which are the left-and right-hand turn probabilities, respectively).

\section {Aharonov-Bohm oscillations of transport characteristics of the Kagomé network}
\label{sec:AB}

In Fig.~3 of the main text, we show the five shortest paths contributing to an electron wave packet propagating from a position ``i'' to ``f''. The wave packet starts and ends on channels propagating in the same direction, and advances along this direction by a period of the long-range variation during the propagation. These paths are the dominant paths contributing to the Aharonov-Bohm oscillations of the electronic transport against an external magnetic field $ - B\bm {e}_z $ when the decoherence length $\ell $ is comparable to the period of the long-range variation ($\ell\sim\Lambda $). 

We show in Fig.~\ref{fig:scattering} two more sets of paths contributing to the Aharonov-Bohm oscillations where the propagation direction of the electron wave packet is flipped. The wave packet propagates forwards along the initial direction by half a long-range variation period to ``$\mathrm{b}_+$'' or backwards by half a long-range variation period to ``$\mathrm{b}_-$''. Summing over the paths shown in Fig.~\ref{fig:scattering}, both of these sets of paths have the same total amplitude,
\begin{equation}
\psi \approx
i e ^ {i\phi/4\phi_0} \sqrt {P_LP_R ^ 2} z ^ 3
+ i e ^ {i\phi/2\phi_0}\sqrt {P_LP_R ^ 5}z ^ 6
-2 i e ^ {i\phi/2\phi_0}\sqrt {P_L ^ 3P_R ^ 3}z ^ 6
+ i e ^ {- i 3\phi/2\phi_0}\sqrt {P_L ^ 5P_R}z ^ 6,
\end{equation}
for $\ell\sim\Lambda $, where $z = e ^ {i \eta / 3} e ^ {i\epsilon\Lambda/2\hbar \mathcal {V}} e ^ {-\Lambda / 4 \ell} $ in terms of the energy $\epsilon $. The magnetic flux through a unit cell of the long-range variation is $\phi = B\mathcal {A} $ (the area of the unit cell is $\mathcal {A} =\sqrt {3}\Lambda ^ 2/2 $) and the magnetic flux quantum is $\phi_0 = h/| e |$. The terms in the amplitude $\psi $ correspond to the amplitudes of paths I, II, IIIa/b (with equal contributions) and IV, respectively. The amplitude $\psi $ exhibits Shubnikov-de Haas style oscillations in the energy $\epsilon $ and Aharonov-Bohm oscillations in the flux $\phi $. As for ``i'' to ``f'', the thermal averaging at high temperatures ($k_B T\gg \hbar \mathcal {V}/\Lambda $ in the main text) suppresses the interference of paths of different length, and only the Aharonov-Bohm oscillations of period $\phi_0 $ commensurate with the long-range variation remain.

At high temperatures, the Aharonov-Bohm oscillations will only feature harmonics in the flux $\phi $ of period an integer multiple of $\phi_0 $ and commensurate with the long-range variation, which we now prove. At such temperatures, only paths of the same length will interfere in the amplitude between two points. The chirality of the channels ensures that the shortest path between the points is unique. We can lengthen the path at each node in one of two ways:
\begin{enumerate}
    \item Add a clockwise loop around a $\mathcal {Q} = 0 $ triangle, increasing the length by $ 3\Lambda/2 $ and the Peierls phase by $ \pi\phi/4\phi_0 $, converting path I to paths II or III. 
    \item Add a counter-clockwise loop around a $\mathcal {Q} = 1 $ hexagon, increasing the length by $ 3\Lambda $ and the Peierls phase by $ - 3\pi\phi/2\phi_0 $.
\end{enumerate}
If the intermediate paths remain connected, then we can add loops and then remove others to create new paths. For example, we convert path I to path IV in Fig.~\ref{fig:scattering} by adding a hexagon and removing a triangle, increasing the length by $ 3\Lambda/2 $ and the Peierls phase by $ -7\pi\phi/4\phi_0 $. Hence, each time the paths grow $ 3\Lambda/2 $ longer, their Peierls phases increase by $ \pi\phi/4\phi_0 $ or $ -7\pi\phi/4\phi_0 $ and the Peierls phase difference is always an integer multiple of $ 2\pi\phi/\phi_0 $.

The paths shown, from ``i'' to ``f'' and ``$\mathrm{b}_\pm$'', are the leading contributions to the Aharonov-Bohm oscillations at high temperatures. This is because they feature the shortest paths (I) where it is possible to add a hexagon then remove a triangle, increasing the Peierls phase by $ -7\pi\phi/4\phi_0 $. This gives the Peierls phase difference of $ 2\pi\phi/\phi_0 $ with the paths where a triangle was added, which give the harmonic of period $\phi_0 $ which dominates the Aharonov-Bohm oscillations.

\bibliography{ref}